# Spatially resolved Raman spectroscopy study of uniform and tapered InAs micro-nano wires: Correlation of strain and polytypism

Vandna Kumari Gupta,[1,2*] Alka A. Ingale,[1,2] Suparna Pal,[1,2] R. Aggarwal[1] and V. Sathe[3]

[1]*Laser Physics Application Section, Raja Ramanna Centre for Advanced Technology, Indore 452013, India.* E-mail address: vandana.gudia8@gmail.com

[2]*Homi Bhabha National Institute, Raja Ramanna Centre for Advanced Technology, Indore 452013, India*

[3]*UGC-DAE Consortium for Scientific Research, Indore 452010, India*

The asymmetric peak ~ 212 - 218 cm$^{-1}$ occuring in InAs micro-nanowires (MNWs: diameter: 2 μm – 300 nm) is investigated using spatially resolved Raman spectroscopy (SRRS) of uniform, bent and long tapered MNWs grown on a Si (001) substrate. It is attributed to superposition of $E_{2h}$ phonon (wurtzite: WZ) and TO phonon (zinc blende: ZB) of InAs. Polarized and wavelength dependent SRRS establishes the presence of WZ and ZB phases in these MNWs. However, formation of WZ phase for larger diameter InAs MNWs is not commensurate with existing growth mapping studies, which needs to be understood further. Study of several of these MNWs suggest that the fraction of WZ to ZB in a MNW is decided not only by diameter, but also by local growth seeding/conditions leading to either tapered or uniform MNWs formation, although, external growth conditions are same. Variation of these frequencies that from bulk value are correlated to residual stress generated in ZB and WZ phases due to presence of WZ and ZB phases, respectively. Consistently, temperature dependent Raman data shows that there is a measurable contribution of stress to dω/dT, a positive for ZB and negative for WZ phonons, due to difference in their thermal expansions. Further, effective thermal expansion coefficient of WZ InAs in presence of ZB phase is calculated to vary in the range 10 - 19 x10$^{-6}$/ K from base to tip of a MNW at ~ 80 K, which is not possible to determine otherwise.

1. **Introduction**

InAs micro-nanowires (MNWs) have potential to play an important role in high-speed optoelectronic material because of it's narrow band gap and high electron mobility [1-5]. The III-V semiconductors e.g. InAs, InSb, GaAs, GaSb, etc. have zinc blende (ZB), whereas, nitrides have wurtzite (WZ) as a stable structure in their bulk form. However, in nanowires (NWs) of these materials, it is found that they can crystallize in both ZB and WZ structures [6-19]. Since, WZ and ZB structure of the same material are expected to have different optoelectronic properties, it is important to understand various manifestations of the polytypism. Raman spectroscopy, in addition to being sensitive to composition, crystalline quality, electronic band structure, size and aspect ratio of NWs; can give information about crystal structure through application of Raman selection rules in different Raman configurations. Raman spectroscopy is especially useful for investigation, wherein density of MNWs is low and is difficult to study using more conventional techniques, like X-ray diffraction (XRD). In our earlier study of laser power dependent time evolution of Raman spectra of InAs MNW (diameter (d) ~ 1 μm), a TO like mode ~ 214 cm$^{-1}$ [20] was noted, whereas, for bulk InAs, TO phonon frequency is ~ 217.5 cm$^{-1}$ [21]. In the present study, we have investigated this observed red shift in InAs MNW, using spatially resolved Raman spectroscopy (SRRS) on several MNWs of different kind. This is an attempt to answer a question, whether, the origin of the red shift lies in polytypism and if so, what is the correlation? In particular, for InAs and GaAs NWs polytypism have been studied by several groups using different techniques [6-8]. Dependence of polytypism on NW diameter i.e. change of phase from ZB to WZ with reducing diameter has been revealed by some recent studies [9-12]. However, it may be important to note that all these reported work are for uniform diameter NWs in the range of 80-200 nm [7-12].

Unlike most of the reported work, where NWs are grown using Au catalyst [9, 11-13,22-25], InAs MNWs study presented here are grown using self-catalyst (In droplet) Metal organic vapour phase epitaxy (MOVPE) [26]. Further, it is important to note that we have studied only one sample in which, we have observed MNWs of diameter varying between 2.2 μm to 200 nm and length varying between 20 - 80 μm. Uniform, tapered and tapered bent MNWs are formed in one single Si (001) substrate with very low density [26]. Thus, first there is a need to validate the study of small diameter (80-200 nm) NWs for these MNWs, as larger diameter MNWs are not expected to have pure ZB structure and then to investigate the effect of polytypism [12,23,27,28]. This is achieved using SRRS along the length of these MNWs, which shows variation in frequency of InAs phonons. Different possibilities are considered, including polytypism for understanding this variation, while establishing the polytypism in these MNWs.

In the second section, experimental details of the study are described. In the third section, results of SRRS and possible origins for the same are discussed. Further, importance of an identification of presence of both ZB and WZ phases in these MNWs using polarized and Resonance Raman spectroscopy is described. The correlation of the study on various types of MNWs is presented, which suggests that sudden change in diameter of wire along the length gives rise to development of large stress leading to bent in MNWs. Stress generated due to presence of both ZB and WZ phases is further investigated using temperature dependent SRRS. In the last section, summary of the work is presented.

## 2. Experimental

The InAs MNWs are grown using self-catalyst MOVPE on Si (001) [26]. Under this growth condition, 10 - 80 µm long MNWs with very low density were formed on the substrate. Unpolarized and polarized Raman measurements were performed using Acton 2500i (single) monochromator with air cooled CCD detector, a part of SPM_integrated Raman system set up, WiTec (Germany) in backscattering geometry at room temperature. Ar ion laser line 488 nm (spectral resolution ~ 3.5 cm$^{-1}$) and He-Cd laser line 441.6 nm with 50x (spatial resolution ~ 1 µm) were used for Raman spectroscopy measurements. The video (optical) image of the MNW was observed before as well as after carrying out the Raman measurement at each individual position on the MNW. Repositioning at same chosen locations at each temperature were guided by visual marks and straight translational movement i.e. care was taken to take data on the same spot for all temperature range using visual guidance. Temperature dependent Raman spectroscopy is performed in the range of 300 K-80 K using liquid N$_2$ and commercial LINKAM THMS 600 stage coupled with LABRAM Jobin-Yvon spectrometer, using 488 nm laser excitation (spectral resolution ~1.5 cm$^{-1}$).

## 3. Results and discussion

InAs MNWs of various dimensions, morphology and tapering factor grown on Si (001) substrate under the same external growth conditions are studied using SRRS.

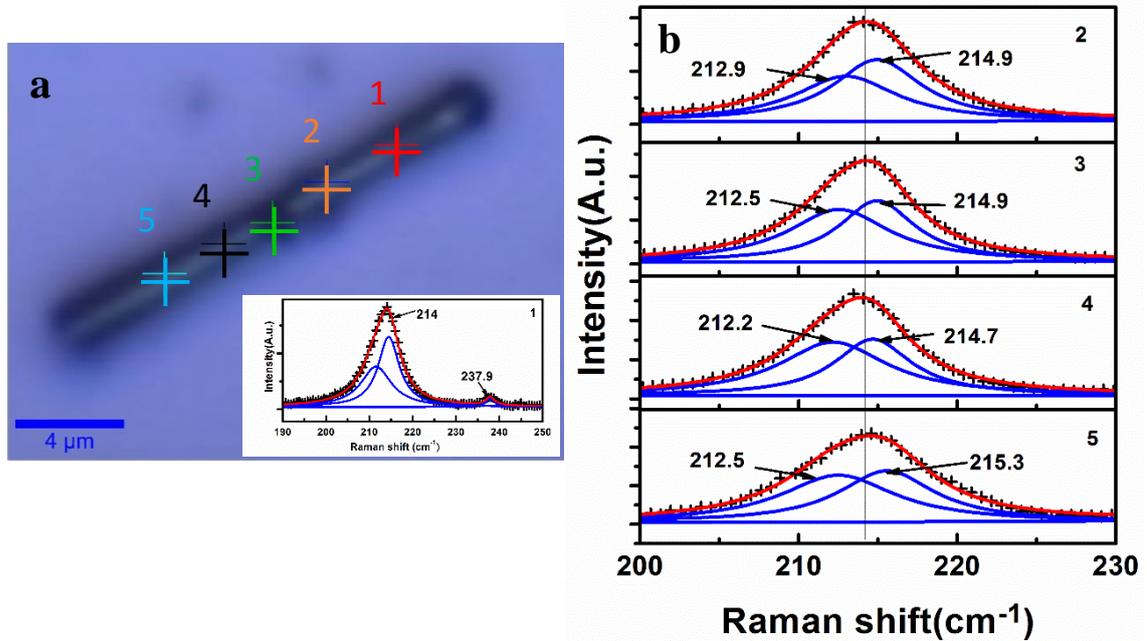

*FIG 1: a) Optical image of uniform InAs MNW (dia ~ 1.2 μm) and b) SRR spectra at positions (2-5) as marked in the optical image. Inset of fig. 1a shows Raman spectrum for position 1 marked in the optical image. Cumulative fit (red solid line) to the raw data ( + ) and separate Lorentzian fits are shown with blue solid line.*

Representative SRR spectra (Fig. 1) for uniform wire (d ~1.2 μm) show strong mode ~ 214 cm$^{-1}$, which can be said to remain constant over the length within experimental accuracy. For various uniform wires studied, frequency of this strong mode varies from ~ 214 cm$^{-1}$ (d ~ 1200 nm) to ~ 213 cm$^{-1}$ (d ~ 600 nm) for different diameters along with the change in lineshape. We have also studied several bent tapered MNWs. Representative Raman data for sharp and smooth bent MNWs is shown in Fig. 2 b and d, respectively. It is important to note that Raman spectra for spatial positions from base to tip show an opposite behavior i.e. in sharp bent MNW, strong mode red shifts as MNW reduces in diameter, whereas, it blue shifts for smooth bent MNW. Before, we discuss origin of these red and blue shifts, it may be interesting to see Raman spectra of long straight tapered MNWs.

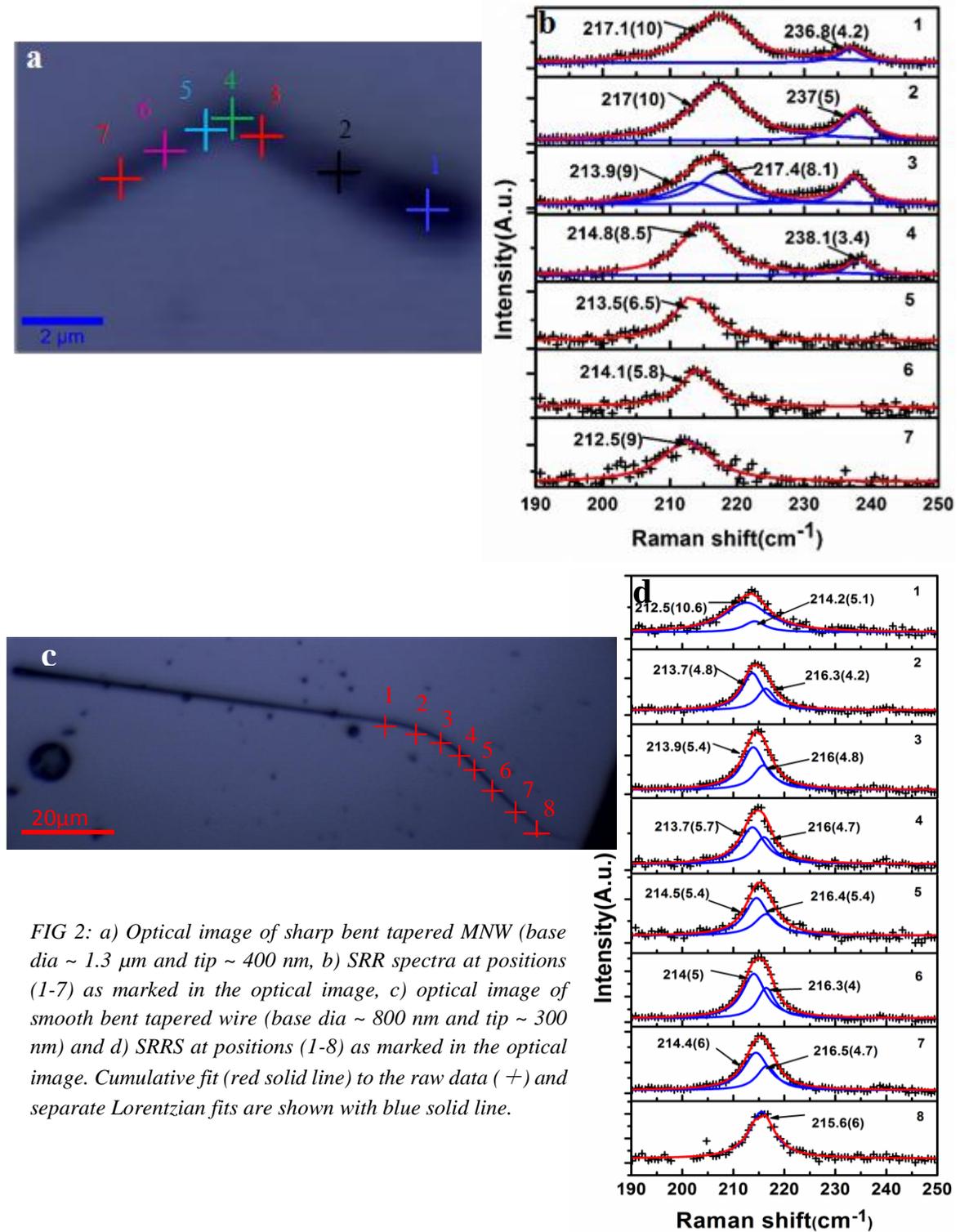

FIG 2: a) Optical image of sharp bent tapered MNW (base dia ~ 1.3 μm and tip ~ 400 nm, b) SRR spectra at positions (1-7) as marked in the optical image, c) optical image of smooth bent tapered wire (base dia ~ 800 nm and tip ~ 300 nm) and d) SRRS at positions (1-8) as marked in the optical image. Cumulative fit (red solid line) to the raw data ( + ) and separate Lorentzian fits are shown with blue solid line.

The SRRS data on many such tapered MNWs shows that in all tapered MNWs bent or otherwise, maximum two peaks are observed in Raman spectra, one strong ~212-218 cm$^{-1}$ (TO phonon like) and sometimes a weak peak ~ 238-239 cm$^{-1}$ (LO phonon). Mode ~212-218 cm$^{-1}$ is said to be TO phonon like as it shows large amount of variation in frequency and lineshape, which is one of the major part of the study presented here. In our case, the full width at half maximum (FWHM) of the TO and LO phonon are found to be ~ 7 cm$^{-1}$ and 4-5 cm$^{-1}$, respectively, indicating a good crystalline quality of the MNWs. For one such long tapered MNW, where, the diameter is ~2 μm at the base (as seen visually) and that at the tip is ~ 400 nm, SRRS data is shown in Fig. 3. A clear monotonic red shift from ~ 218 cm$^{-1}$ to ~212 cm$^{-1}$ is observed in the TO like phonon frequency as we move from the broader base region to thinner tip region. It is important to note that line shape is asymmetric and changes along the length of MNWs as shown in Fig. 3. The Raman data of points 7 and 8 are not plotted as signal to noise ratio is poor; however, it shows further red shift.

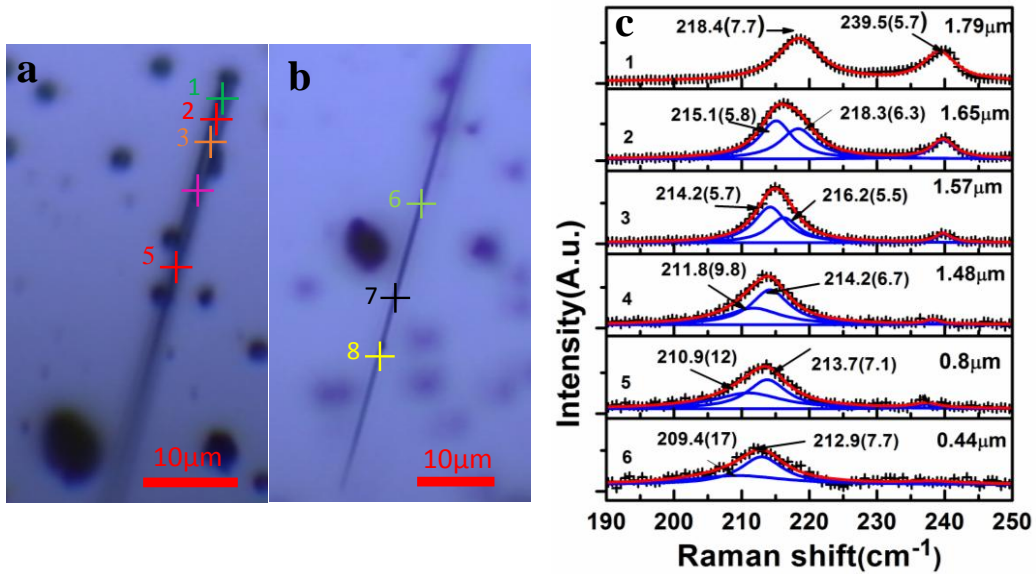

*FIG 3: a) Optical image of straight tapered MNW, wherein objective is focused near the base (dia ~ 2 μm), b) Optical image of same wire, wherein objective is focused at the tip (~ 400 nm) and c) SRRS at positions (1-6) as marked in both optical images. Cumulative fit (red solid line) to the raw data (+) and separate Lorentzian fits are shown with blue solid line.*

SRR spectra of dots show much stronger LO phonon as shown in Fig. 4. Here, LO and TO modes can be fitted using Lorentzian and both the frequencies obtained are slightly blue shifted from bulk. This may be due to residual compressive strain. The observation of both TO and LO phonons in backscattering geometry suggests non-epitaxial growth on Si (001) substrate. This is because from (001) phase of cubic InAs, only LO phonons are allowed in backscattering geometry. With this background, the origin of red and blue shift in MNWs is explored in the following. The peak ~235 cm$^{-1}$ may be a L$_-$ coupled plasmon-phonon (LO) mode due to photoexcited carriers. This will be investigated separately and is out of scope of this work.

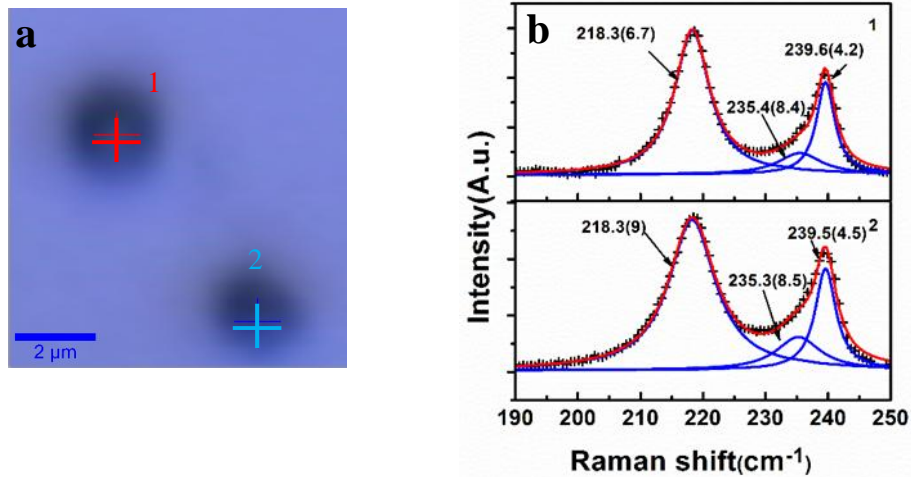

*FIG 4: a) Optical image of InAs micro sphere and b) SRRS at positions marked in fig. 4a. Cumulative fit (red solid line) to the raw data ( + ) and separate Lorentzian fits are shown with blue solid line.*

The red shift observed in long straight uniform and tapered MNWs can have three possible origins, i) Heating due to laser irradiation, ii) Confinement effect and iii) residual stress. The Raman data is taken with 488 nm with laser power ~ 0.3 mW. Heating effect is discarded as a possible origin of red shift from our earlier laser power dependent time evolution Raman study performed on uniform wire (d ~ 1 μm), which shows blue shift and reduction in FWHM at higher powers (6 mW), whereas, larger redshift and increase in FWHM is expected, if it was due to heating [20]. The effect of confinement is also rejected, since diameters we are studying are much beyond phonon confinement effects are expected to occur (< 200nm). Therefore, we consider stress to be the cause of red as well as blue shift observed in these MNWs. The first possibility to be considered for generation of stress [6, 13] is effect of polytypism observed in III-V NWs grown under various conditions [6-19]. In bulk form, all III-V semiconductors except nitrides have ZB structure, however, recent studies shows that III-V NWs grow with polytypism i.e. alternate ZB and WZ structure. This is studied in detail by some groups recently using different techniques [6-8]. The ratio of WZ to ZB content of these structures depends on various growth parameters such as growth temperature, metal nanoparticle diameter, V/III ratio, total mass flow on crystal structure etc. Most authors have reported [9,11,12] that the crossover diameter graph becomes broader and shifts to larger diameter as growth temperature changes from 480 to 420 $^0$C for V/III ratio ~ 130 with Au-catalyst method. Increasing in growth temperature has favoured ZB and WZ depending on the range of temperature as well as V/III ratio [10, 12, 13, 19, 27]. In totality, growth mapping studies especially for InAs NWs suggest that change in ZB to WZ content is not necessarily a smooth function of change in growth parameters. It may be important to note that very small change of V/III ratio/temperature/mass flow may significantly change crystal structure content, whereas, over a large change of these growth parameters, their hardly any change in the content [19 and references therein]. Although, several such studies performed using V/III ratio (varying from 20-220) and temperature (varying from 380 to 500$^0$C) suggest that our growth conditions (growth temperature ~ 425 $^0$C and V/III ratio ~ 250) may lead to mainly ZB structure. However, since the data presented above suggest otherwise, it needs to be further confirmed. That is to say, although, we consider the reason of red shift to be stress, however, with the above background, whether, the reason of stress lies in polytypism or not, needs to be further investigated in our case. Spatially resolved polarized Raman spectroscopy need to be performed in order to obtain information on the polytypism. Before performing polarized Raman spectra, we analyze Raman data considering this possibility to understand the trend in these MNWs.

For the two possible structures ZB and WZ, number of atoms per unit cell is 2 and 4, respectively. Thus, total number of optical phonon modes is 3 and 9 in ZB and WZ, respectively. Out of these, $A_1$(TO), $E_1$(TO), $A_1$(LO) and $E_1$(LO) modes lie very close (within ~ 0.5 cm$^{-1}$) in phonon frequencies for WZ, which are also close to TO and LO phonons for ZB structure [22]. However, difference between $E_{2h}$ and TO phonon modes is expected to be ~ 6.3 cm$^{-1}$ for WZ InAs [22] and this, therefore can be used to get information about the presence of ZB and WZ phases in InAs MNWs. Considering the possibility of presence of $E_{2h}$ and TO phonon in TO like phonon noted above, all the spectra of Fig.1-4 are deconvoluted with minimum required Lorentzian i.e. either two or three. Upon deconvolution, two modes; lower frequency ($\omega_l$) and higher frequency ($\omega_h$) are observed. For the uniform diameter wire, we observe $\omega_l$ and $\omega_h$ to be ~212.5 cm$^{-1}$, ~214.5 cm$^{-1}$ for d ~ 1200 nm, throughout the length of the wire. However, for tapered long MNW (Fig 3), $\omega_h$ and $\omega_l$ varies from ~ 218 and 215 cm$^{-1}$ at the base (d~1.6 μm) to the tip (d ~ 400 nm), where $\omega_h$ disappears and $\omega_l$ reaches the value of ~213 cm$^{-1}$. It is interesting to observe that as we progress from position 1 to 2, $\omega_l$ mode is appearing and at further position 3, the intensity of $\omega_h$ mode is decreasing. The $\omega_l$ and $\omega_h$ is red shifted to ~ 216.2 cm$^{-1}$ and 214.2 cm$^{-1}$, respectively. At position 4 to 6 where the diameter reaches a value ~ 400 nm, $\omega_h$ disappears and $\omega_l$ dominates till position 6, one new mode starts appearing ~ 211 - 209 cm$^{-1}$ (FWHM increases from position 4 to

6), when $\omega_l$ reaches a frequency value of ~213 cm$^{-1}$. As one can see, $\omega_l$ and $\omega_h$ observed here are very close in frequency (within ~2 cm$^{-1}$) and depending on the fraction of ZB/WZ the identification of $\omega_l$ and $\omega_h$ as $E_{2h}$ and TO (ZB + WZ/ZB) may or may not be correct, specifically, as it's not feasible to have any supportive information via more conventional techniques like transmission electron microscopy (TEM) or XRD in this case. Therefore, before we discuss the generation of stress due to presence of ZB and WZ phases in these InAs MNWs, we perform polarization dependent Raman measurements to identify the origin of phonon frequencies obtained on deconvolution.

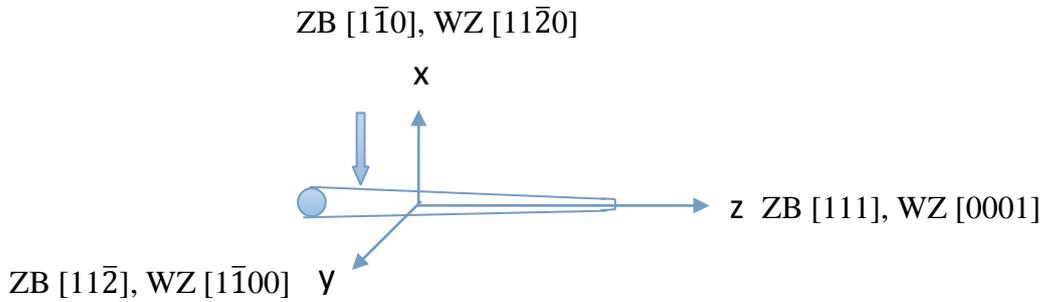

FIG 5: Schematic diagram of the Raman scattering configuration w.r.t. NW axis taken to be in Z direction for back scattering geometry. x and $\bar{x}$ axes are the incident and scattering polarization. Z axis is taken to be [111] and [0001] for ZB and WZ phases, respectively.

The schematic of geometry of polarized Raman measurement is shown in Fig. 5. The z direction is chosen parallel to growth axis of MNW, which is (111) for ZB and (0001) for WZ structure. It is well established that NW growth happens along (111) ZB and or (0001) WZ directions [6,7,10,12,22,27,29-30]. Before, we go further, it may be appropriate to discuss the reasoning behind formation of ZB and WZ phases in InAs NWs. The atomic arrangements in these two structures are very close to each other along the above mentioned directions except for the azimuthal rotation [31, 32]. In addition, difference between minimum energy for ZB and WZ phases is calculated to be very small in the range -18 to -9 meV/atom for III-N, 3 to 12 meV/atom for III-V and -1.1 to 6 meV/atom for II-VI group [10, 31, 33]. Therefore, existence of both phases for II-VI and III-N can be considered to be a finite possibility and has been already observed for InN [34-36] and CdS [37-40] in bulk form. However, most III-V's except some nitrides shows cubic structure in bulk form. However, NW of III-V's like GaAs, InAs stablizess in pure WZ, ZB and mixed phases depending on growth conditions [6-11,22,27,28,30,31]. Volker et al. have reported that the bulk energy per atom pair for ZB phase is lower than WZ bulk energy, therefore, the stable phase is ZB in bulk material in this case [10], however, for NWs, the WZ phase has lower surface energies than ZB phase for corresponding crystalline orientations in the same material. This effect is expected to stabilize the WZ structure for sufficiently thin NW, that is, when the surface-to-volume ratio is high enough [9-10, 31]. However, in our case, where diameters range from 2 μm to 300 nm, this need not be the case. Therefore, it is very important to do polarization studies to first confirm the possibility of existence of both the ZB and WZ in studied MNWs.

*TABLE I: Raman selection rules for different scattering geometries [6] are summarized.*

| | |
|---|---|
| x(z,z)x̄ | TO : ZB & $A_1$(TO) :WZ |
| x(y,y)x̄ | TO : ZB , $A_1$(TO) : WZ &  $E_{2h}$:WZ |

Z direction is considered to be MNW growth direction. For cubic structure, x, y are chosen to be direction [1$\bar{1}$0], [11$\bar{2}$]; perpendicular to growth directions [6-8, 28]. The back scattering is taken to be along x [1$\bar{1}$0] and [11$\bar{2}$] direction. We have performed Raman measurement on a single uniform MNW with a laser light 488 nm in two polarization configuration x(z,z)x̄ and x(y,y)x̄ i.e. incident as well as scattered light is polarized i) parallel and ii) perpendicular to MNW axis, respectively. Raman selection rules suggests that the TO phonon of ZB/WZ and $E_{2h}$ of WZ are allowed in $x(y,y)\bar{x}$ geometry, and only TO phonons of ZB/WZ are allowed in $x(z,z)\bar{x}$ geometry (Table I) [6]. Thus, we find that study on these two geometries itself can give required information regarding WZ and ZB phase for InAs MNWs. The Representative polarized Raman spectra for uniform wire of diameter ~ 600 nm is shown in Fig.6.

Several such data is repeated at various sites on several MNWs. All spectra show two main structures, one strong peak ~212-217 cm$^{-1}$ and other weaker peak ~238-239 cm$^{-1}$. The first is deconvoluted into two modes corresponding to $\omega_l$ and $\omega_h$ phonons. The $\omega_l$ lies in the range ~ 212-214 cm$^{-1}$, whereas, $\omega_h$ lies in the range ~ 215-218 cm$^{-1}$. However, for some MNWs, additional broad mode with FWHM >10 cm$^{-1}$ can be deconvoluted from this peak along with $\omega_l$ in the range of 206-209 cm$^{-1}$ as noted earlier. This is assigned as silent $B_1$ mode for WZ phase of InAs, which is being allowed due to disorder. Fig. 6 shows polarization Raman data for uniform wire (d ~ 800 nm). Polarization data is repeated several times on different uniform wires to confirm the above noted observations.

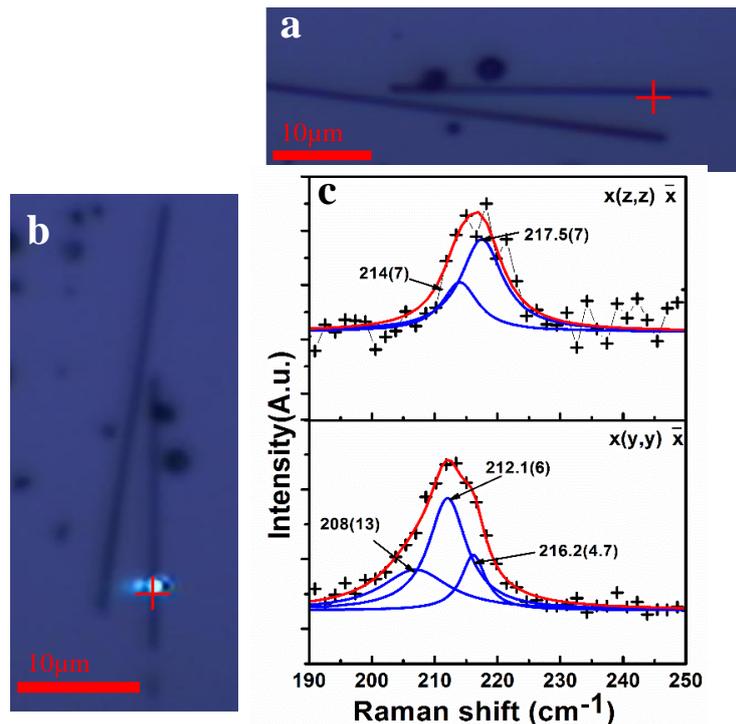

*FIG 6: Optical image of uniform wire (dia ~ 600 nm) a) in horizontal direction, b) in vertical direction and c) Polarized Raman data in $x(z,z)\bar{x}$ and $x(y,y)\bar{x}$ configuration at positions as marked in optical image (a) and (b), respectively. Cumulative fit (red solid line) to the raw data ( + ) and separate Lorentzian compartments are shown with blue solid line.*

Further, the unpolarized data is found to replicate Raman data in **x(y,y)x̄** configuration probably due it's large signal in this configuration. The observation of red shift of TO like mode in x(y,y)x̄ configuration compared to that of x(z,z)x̄ configuration, where $E_{2h}$ is expected to dominate the spectra clearly indicates that $\omega_l$ is indeed $E_{2h}$. Similar observations are reported earlier for III-V NWs in the range ~100 nm [6,7,22].

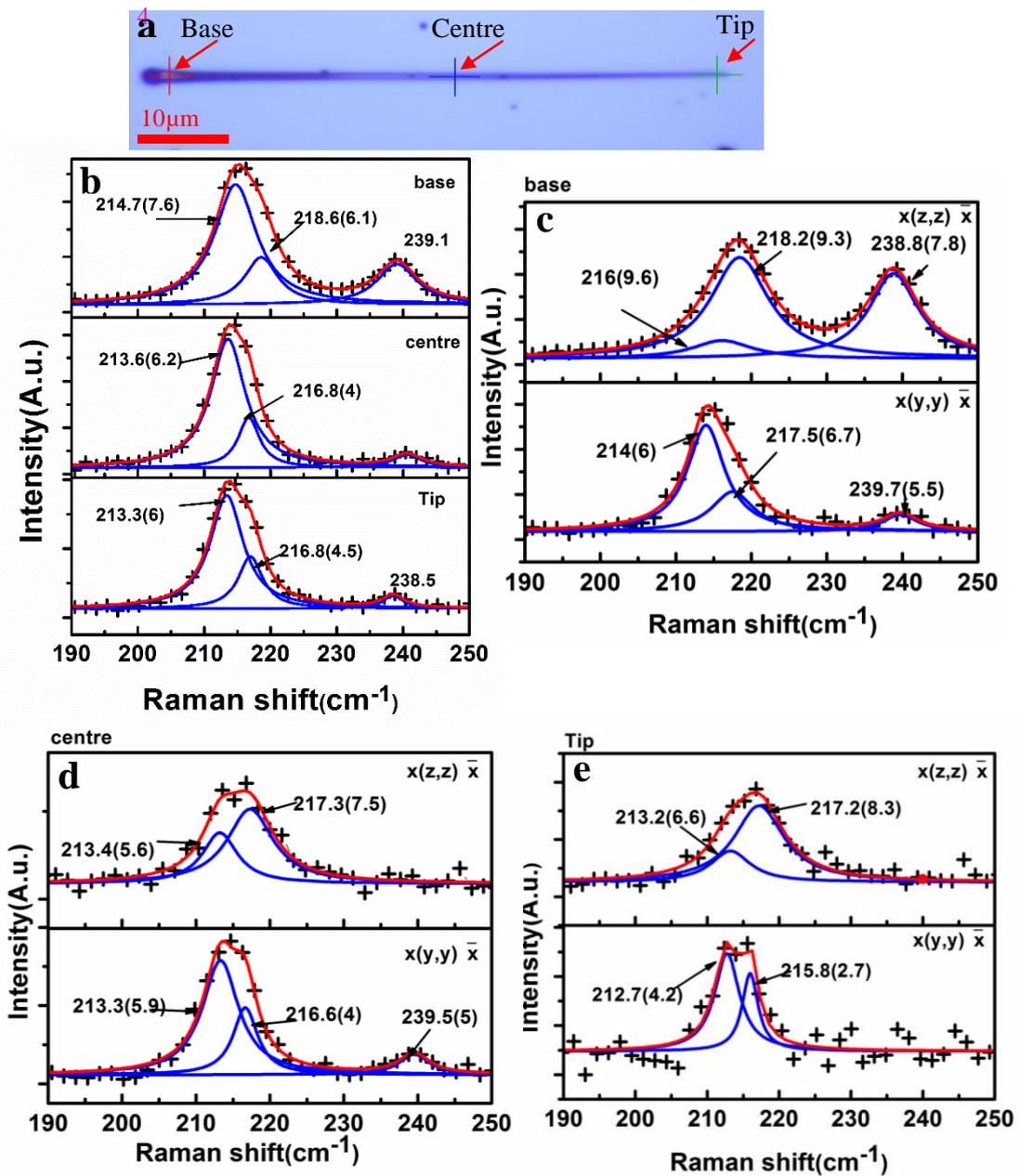

FIG 7: a) Optical image shows tapered MNW (base dia ~ 2 μm and center dia ~ 1.6 μm and tip ~ 800 nm), b) unpolarized Raman data at position marked for base, center and tip, c), d) and e) Polarized Raman data for base center and tip position, respectively in x(z,z)x̄ and x(y,y)x̄ configuration. Cumulative fit (red solid line) to the raw data (+) and separate Lorentzian fits are shown with blue solid line.

We also performed polarization dependent Raman measurement on tapered MNWs too. The unpolarized and polarized spectra are shown in Fig.7 for base, center and tip. In polarized Raman spectra, $\omega_l$ and $\omega_h$ varies from ~216 to ~213 cm$^{-1}$ and ~218 to ~217 cm$^{-1}$ in $x(z,z)\bar{x}$ configuration, whereas, from ~214 to ~ 212.7 cm$^{-1}$ and ~217.5 to ~215.8 cm$^{-1}$ in $x(y,y)\bar{x}$ configuration, respectively for MNW going from base (d ~ 2 μm) to tip (d ~ 800 nm). The line shape and width of these modes ~215 cm$^{-1}$ in $x(y,y)\bar{x}$ geometry suggests that frequencies of $\omega_l$ to $\omega_h$ comes closer continuously from base to tip. The relative intensity ratio however does not seem to change significantly. The presence of $E_{2h}$ phonon with very small intensity in $x(z,z)\bar{x}$ geometry may be due to disorder. This is discussed in next section. Further, as for uniform diameter MNWs, it is also found that unpolarized data is replica of $x(y,y)\bar{x}$ data in frequency and lineshape. In Fig. 8, we have plotted $\omega_l$ and $\omega_h$ v/s diameter for uniform and tapered MNWs deconvoluted from unpolarized data. It is important to note here that FWHM decreases for both TO and $E_{2h}$ phonons in $x(y,y)\bar{x}$ Raman scattering configuration, which clearly brings out the fact that crystalline quality of the material in the direction of NW axis is poorer than perpendicular to it. This is easy to understand as there are many stacking faults due to ZB and WZ structure along the length of MNW axis, whereas, crystalline quality is much better perpendicular to the MNW axis. As lateral dimension decreases i.e. as we go from base to tip, overall crystalline quality increases.

*TABLE II: Diameter of base, center and tip positions for tapered wires (TW) designated as 1, 2, 3, 4, 5 and 6 for which phonon frequencies are plotted in Fig. 8. L: Length of N/MW, D: diameter.*

|  | Base D(μm) | Centre D(μm) | Tip D(μm) |
| --- | --- | --- | --- |
| TW 1(L -80 μm) | 1.57 | 0.78 | 0.4 |
| TW 2 (L- 24 μm) | 2.2 | 1.93 | 0.717 |
| TW 3 (L- 61 μm) | 1.77 | 0.824 | 0.588 |
| TW 4(L- 49 μm) | 1.76 | 1.17 | 0.823 |
| TW 5(L- 32 μm) | 2 | 1.46 | 0.8 |
| TW 6(L- 62 μm) | 2.1 | 1.6 | 0.78 |

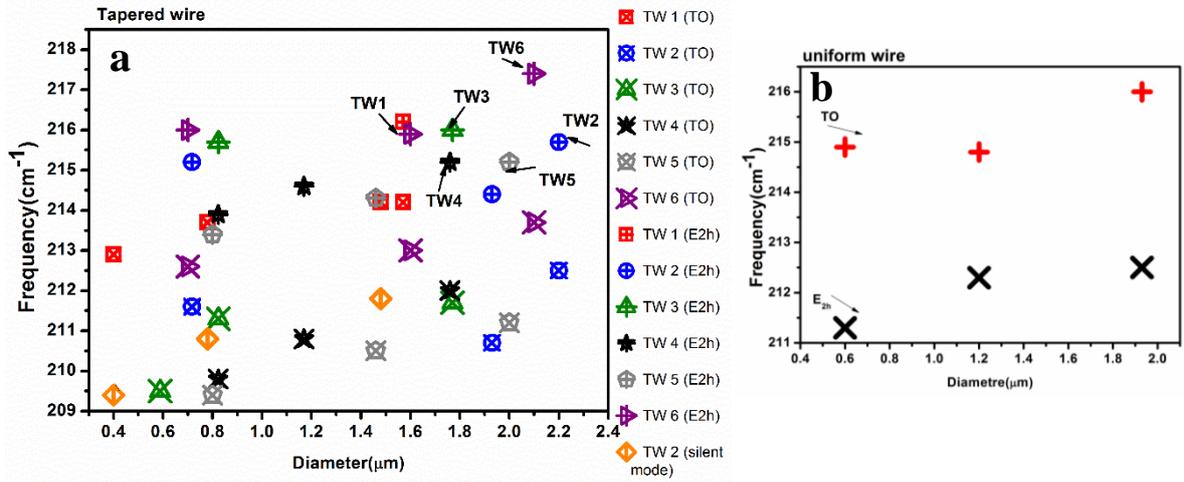

*FIG 8: a) shows diameter v/s frequency for 6 Tapered wire (TW). Here plus (+) sign with red, blue, green, black, gray, dark pink colors show TO of ZB for TW 1,2,3,4,5,6, respectively. Cross (X) with red, blue, green, black, gray, dark pink colors show $E_{2h}$ of WZ for TW 1,2,3,4,5,6, respectively and b) shows diameter v/s frequency for 3 uniform wires. Symbols (+) and (X) show TO and $E_{2h}$, respectively.*

The $\omega_l$ and $\omega_h$ thus obtained are plotted for corresponding diameters for several tapered and uniform MNWs separately in Fig. 8. It is interesting to note that the variation observed in value of frequency for $\omega_l$ and $\omega_h$ is quite large, especially for $\omega_l$. Further, it is found that this depends not only on diameter, but also on it's position on the tapered MNW i.e. if the diameter measured is at the base or center or tip of the MNW. This indicates that the residual strain varies with diameter and reason behind it needs to be understood for MNWs. In the following, we discuss these results in the light of polytypism observed in these InAs MNWs.

**3.1. Effect of polytypism on Raman spectra**

The preferred growth direction of III-V NWs has been known to be (111) cubic and for II-VI NWs is known to be (0001) for mixed phase growth in III-V NWs [6,7, 22,31]. It is also known that (111) cubic and (0001) hexagonal are equivalent directions i.e. atomic placement in the two directions for two types of atoms is very similar and can go from one to another with slight azimuthal rotation and therefore for NWs such change of phase is observed [32]. When such stacking fault of change in phase from ZB to WZ happens, it is expected to manifest itself in two ways, i) breaking of selection rules and or ii) generation of strain in the ZB and WZ due to presence of the other. The TO phonon frequency of bulk InAs is found to be ~ 217.5 cm$^{-1}$ [21,7,28,41,42] and $E_{2h}$ of WZ is calculated to be ~ 211±1cm$^{-1}$[22].

In principal, intensity ratio of $E_{2h}$ and TO phonon may give us some clue about change in WZ/ZB content, considering the earlier reports, which suggests that TO phonon observed is mainly due to ZB phase and the fact that they show change of frequency in different direction from that of the bulk value [6-8,22]. However, Zardo et al. has shown that $E_{2h}$ and TO phonon shows very different resonance profiles ($E_{2h}$ peaks ~2.4 and TO peak ~2.7eV) for Raman scattering for InAs NW due to difference in their structure i.e. WZ and ZB [8, 22, 43, 44] and hence, the ratio may not be commensurate with the content of ZB to WZ phases, when measured with 488/514.5 nm excitations.

In Fig. 9, we show polarized Raman spectra at same position of an InAs MNW for 488 and 441.6 nm excitations. One can see that intensity of LO and TO increases drastically compared to that of structure $E_{2h}$ at 441.6 nm excitation for the geometry, wherein both A1 (TO) and $E_{2h}$ are allowed. The intensity ratio of TO and $E_{2h}$ therefore cannot be guiding parameter for obtaining fraction of ZB to WZ phase. Nevertheless, this clearly establishes that the TO phonon observed in unpolarized data is mainly due to ZB phase. Although, intensity ratio is not a good measure of ZB/WZ ratio, it is noted that as WZ content increases, TO phonon frequency (ZB) shows red shift, indicating tensile stress, whereas, $E_{2h}$ phonon frequency (WZ) shows blue shift, suggesting compressive stress. Similar results have been noted earlier for GaAs and InAs uniform NW of diameter ~100-200 nm [6,7]. To the best of our knowledge, this is the only related study for III-V NWs. In this context, it may be important to note here, that the stress generated is not due to lattice mismatch occurring in an epitaxial growth (ZB phase/WZ phase alternating as substrate and grown layer) as suggested by Zardo et al. [6, 45].

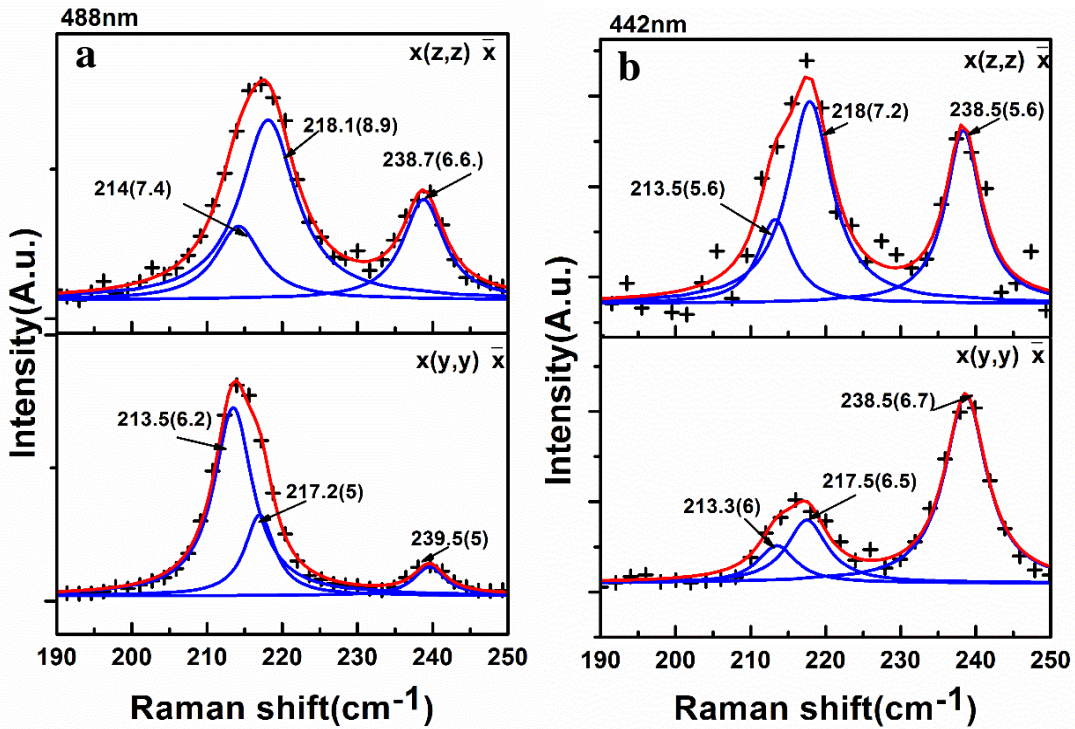

*FIG 9: a) and b) Polarized Raman data performed with 488 nm and 442 nm excitation, respectively at same wire and at similar position in $x(z,z)\bar{x}$ and $x(y,y)\bar{x}$ configuration for uniform wire (dia ~ 1.2 μm).*

Although, lattice parameters for ZB and WZ phases are different, as expected due to geometric reconfiguration of the unit cell, the faces/planes (111 for ZB and 0001 of WZ) have almost same atomic arrangement except azimuthal rotation. That is to say, although, growth is epitaxial in nature in the direction of growth axis, the phase and planes are different, so that atomic arrangements are similar for two cases. However, atomic arrangements differ significantly in other planes. Similar strain generated due to polytypism is studied for (ZB/WZ) layered SiC thin films [46]. Based on various experimental studies, the strain due to polytypism in ZB and WZ structures is correlated to formation of 2H, 4H and 6H, 3C as more stable unit cell structure in case of polytype growth direction, thereby changing lattice constant in the perpendicular direction too. This in turn leads to either compressive or tensile stress, value of which depends on percentage of hexagonality and internal cell parameter [13, 33, 46-51]. Theoretically also, it is found to be minimum energy configuration/s due to changed next near neighbor interactions and therefore, the strain generated is expected to be proportional to content of the other polytype [13]. We attribute the generation of strain due to polytypism in MNWs to be the same and hence they can be used to calculate ratio of content ZB to WZ phase once calibrated to do so. For this, more systematic study along with other technique which can probe this content independently is required, which will be pursued separately and is out of scope of this paper. In the following, however, we can comment upon relative change in ZB to WZ content ratio. The effect of WZ/ZB fraction for uniform and tapered MNWs with different diameter is further studied using temperature dependent Raman spectroscopy measurements. Zardo et al. have studied variation in phonon frequency of $E_{2h}$ (WZ phase) and $E_1$ (TO: ZB) with change in relative percentage of WZ and ZB for a uniform diameter GaAs NW [6].

They have varied percentage of WZ and ZB along the length of NW by changing V/III ratio during the deposition. They have observed that the percentage of WZ varies from 98% at one end of NW to 32% in the center and near 0% at another end of NW by High Resolution TEM measurement. All our observations regarding blue and red shift for tapered and uniform MNWs and understanding developed about WZ/ZB ratio for the same is commensurate with observations made by Zardo et al. for above mentioned GaAs NW [6].

In order to avoid contributions due to all possible variations as noted above, we have performed temperature dependent, polarization dependent Raman measurements and atomic force microscopy measurements on the same tapered MNW for one to one correlation. Since, polarized Raman data and line shapes of unpolarized Raman data clearly suggests presence of two peaks for a structure ~ 212-218 cm$^{-1}$, we have fitted spectra from 80 K to 300 K for base to tip with three Lorentzian function for LO, TO and E$_{2h}$ keeping frequency of the $\omega_{LO}$, $\omega_{TO}$, $\omega_{E2h}$, width ($\Gamma_{LO}$, $\Gamma_{TO}$, $\Gamma_{E2h}$) and intensities of LO, TO, E$_{2h}$ as fitting parameters. Before going further, we shall like to note that due to significant mixing of phases as noticed from the frequencies of E$_{2h}$ and TO phonons, we believe that in our case, calculation of fraction using following equation may not be feasible.

$$\left(\frac{d\omega}{dT}\right)_{Total} = x \left(\frac{d\omega}{dT}\right)_{ZB} + (1-x) \left(\frac{d\omega}{dT}\right)_{WZ} \quad (1)$$

Indeed, we get x = -0.6, an unfeasible value for concentration, with d$\omega$/dT$_{total}$ ~ -0.01450, d$\omega_{E2h}$/dT$_{WZ}$ and d$\omega_{TO}$/dT$_{ZB}$ are taken to ~ -0.005 and -0.01 respectively [42]. This suggests that the contribution of residual stress needs to be exclusively taken into account and secondly phonons both originating from ZB and WZ phases need to be treated separately, as they undergo different kind of stress. Thus equation (1) can be written as, [42].

$$\left(\frac{d\omega}{dT}\right)_{WZ} = \left(\frac{d\omega}{dT}\right)_{TE} + \left(\frac{d\omega}{dT}\right)_{S} \tag{2}$$

$$\left(\frac{d\omega}{dT}\right)_{ZB} = \left(\frac{d\omega}{dT}\right)_{TE} + \left(\frac{d\omega}{dT}\right)_{S} \tag{3}$$

Frequencies v/s temperature of deconvoluted peaks is plotted in Fig. 10 (a, b, c, d, e, f) for base ($P_1$), center ($P_3$) and tip ($P_5$) for the tapered MNW. It may be appropriate to note here that unlike others [41, 42, 8], we are able to resolve peak ~212-218 cm$^{-1}$ into two peaks i.e. TO phonon and $E_{2h}$. This has become feasible as unlike others, we have observed Raman data for larger MWNs, which have larger ZB content. Whereas, earlier reported work is mainly on smaller NWs (~100 nm), wherein, WZ is a dominant phase [7, 8, 22, 42]. The slope of the plots for $E_{2h}$ and TO phonons $d\omega/dT_{E2h}$, $d\omega/dT_{TO}$ and $d\omega/dT_{LO}$ are calculated for base ($P_1$ ~ 2 µm), $P_2$ (~1.7 µm), center ($P_3$ ~ 1.6 µm), $P_4$ (~ 1µm) and tip ($P_5$ ~ 800 nm) and for uniform MNW ($P_8$ ~ 500 nm) are tabulated in table III. One can clearly see that for TO and LO phonons, frequencies are less than that for bulk ZB [42, 52] and for $E_{2h}$, it is greater than that of bulk WZ [42].

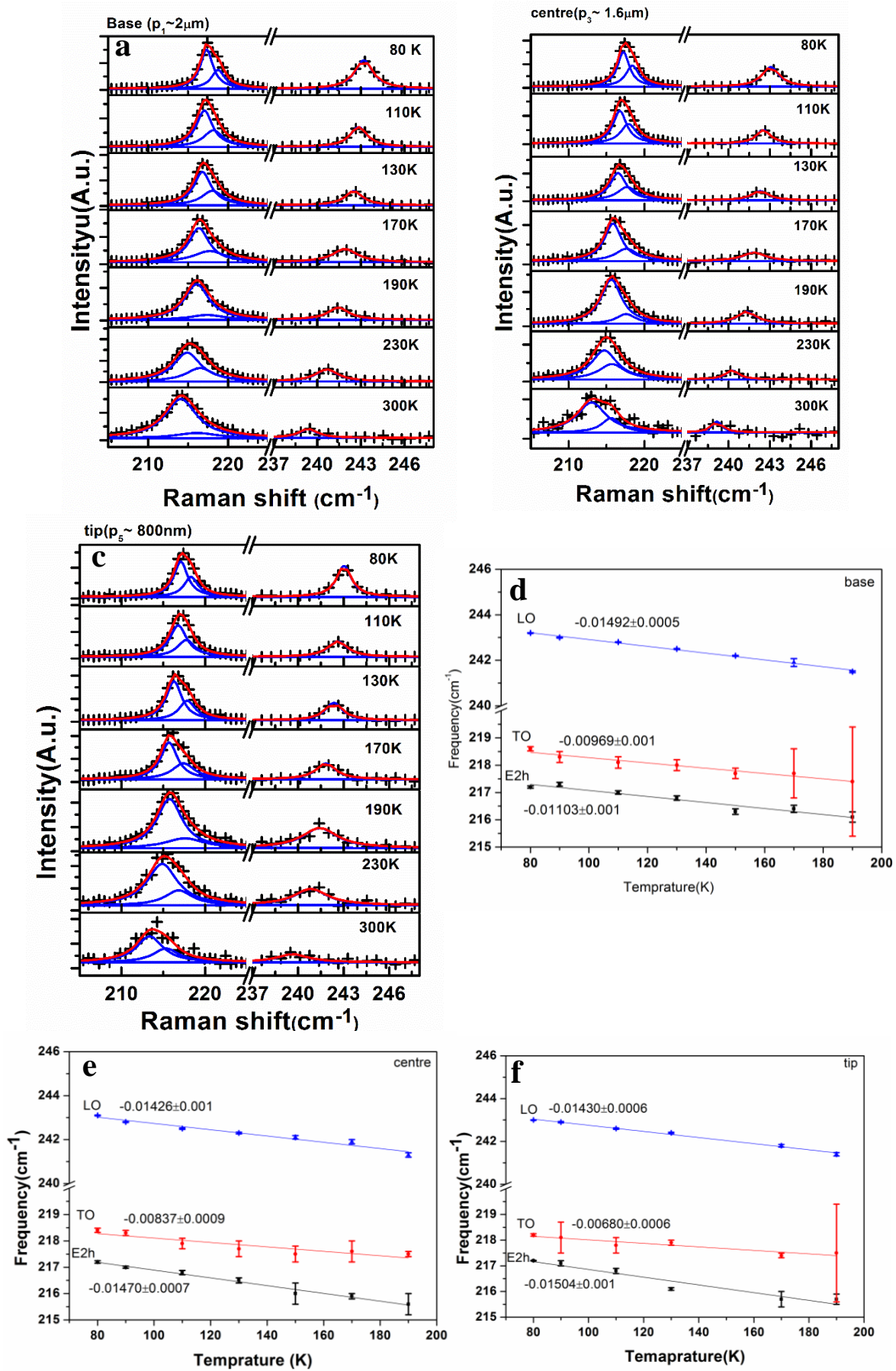

FIG 10: a), b) and c) show the temperature dependent Raman spectra in the temperature range 300K to 80K for base (~2 μm), center (~1.6 μm) and tip (~800 nm) position, respectively. Cumulative fit (red solid line) to the raw data ( +) and separate Lorentzian fits are shown with blue solid line. Frequency v/s temperature plot showing linear fit for TO, LO and $E_{2h}$ frequencies for d) base, e) center and f) tip position. The Error bar corresponds to the standard error to ω as obtained from nonlinear least square fir to the spectra. The solid line are the best fit to the data points, from which slope are calculated in fig. d)-f).

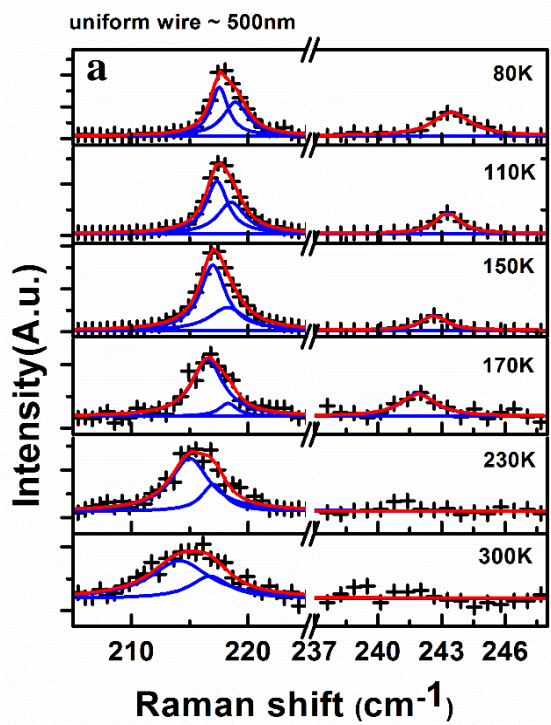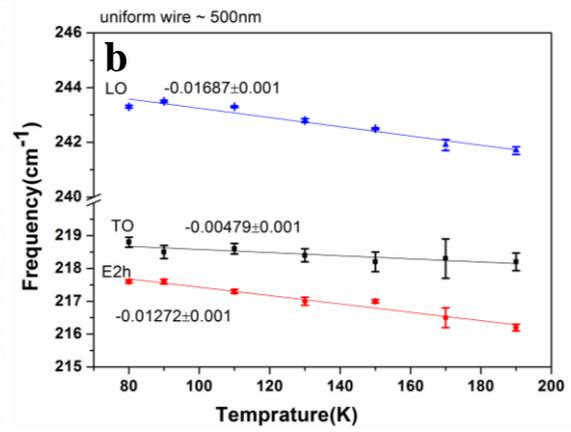

*FIG 11: a) Show the temperature dependent Raman spectra in the temperature range 300K to 80K for uniform wire (dia ~ 500 nm). Cumulative fit (red solid line) to the raw data ( + ) and separate Lorentzian fits are shown with blue solid line and b) Frequency v/s temperature plot showing linear fit for TO, LO and $E_{2h}$ frequencies for uniform wire. The solid lines are the best fit to the data points.*

*TABLE III: Frequencies of $E_{2h}$ ($\omega_l$) and TO ($\omega_h$), intensity ratio of TO to $E_{2h}$ phonons at 300K and 80K and $d\omega_{E2h}/dT$, $d\omega_{TO}/dT$, $d\omega_{LO}/dT$ for different positions on a tapered MNW and for a uniform wire are summarized.*

| | $\omega_l$, $\omega_h$ and ratio of $\omega_h/\omega_l$ at 300K | $\omega_l$, $\omega_h$ and ratio of $\omega_h/\omega_l$ at 80K | $d\omega_{E2h}/dT$ (cm$^{-1}$/K) | $d\omega_{TO}/dT$ (cm$^{-1}$/K) | $d\omega_{LO}/dT$ (cm$^{-1}$/K) |
|---|---|---|---|---|---|
| Base ($P_1 \sim 2$ μm) | 214.1, 216.1 and 0.17 | 217.2, 218.6 and 0.5 | -0.01103 ± 0.001 | **-0.00969 ± 0.001** | -0.01492 ± 0.0005 |
| $P_2 \sim 1.7$ μm | 213.4, 215.4 and 0.35 | 217.4, 218.5 and 0.49 | -0.01412 ± 0.001 | **-0.00955 ± 0.001** | -0.01482 ± 0.0007 |
| Center ($P_3 \sim 1.6$ μm) | 213, 215.6 and 0.31 | 217.2, 218.4 and 0.6 | -0.01470 ± 0.0007 | -0.00837 ± 0.0009 | -0.01426 ± 0.001 |
| $P_4 \sim 1$ μm | 213.2, 215.5 and 0.38 | 217.4, 218.6 and 0.32 | -0.01483 ± 0.001 | -0.00805 ± 0.0008 | -0.01431 ± 0.0008 |
| Tip ($P_5 \sim 800$ nm) | 213.3, 215.3 and 0.52 | 217.1, 218.3 and 0.57 | -0.01504 ± 0.001 | -0.00680 ± 0.0006 | -0.01430 ± 0.0006 |
| uniform wire ($P_8 \sim 500$ nm) | 213.6, 216.5 and 0.75 | 217.5, 218.9 and 0.8 | -0.01272 ± 0.001 | <span style="color:red">**-0.00479 ± 0.001**</span> | -0.01687 ± 0.001 |

TABLE IV: Strain component of $(d\omega_{E2h}/dT)_S$, $(d\omega_{TO}/dT)_S$ and $(d\omega_{LO}/dT)_S$ for different positions on a tapered MNW and for a uniform wire are summarized.

| diameter | $(d\omega_{E2h}/dT)_S$ (cm$^{-1}$/K) | $(d\omega_{TO}/dT)_S$ (cm$^{-1}$/K) | $(d\omega_{LO}/dT)_S$ (cm$^{-1}$/K) |
|---|---|---|---|
| P$_1$ ~ 2 µm (TW: Base) | -0.0060± 0.001 | **+0.0003±0.001** | +0.0020±0.0005 |
| P$_2$ ~ 1.7 µm | -0.0091±0.001 | **+0.0005±0.001** | +0.0022±0.0007 |
| P$_3$ ~ 1.6 µm (TW: Center) | -0.0097± 0.0007 | **+0.0016±0.0009** | +0.0027± 0.001 |
| P$_4$ ~ 1 µm | -0.0098±0.001 | +0.0020±0.0008 | +0.0026±0.0008 |
| P$_5$ ~ 800 nm (TW: Tip) | -0.0100± 0.001 | +0.0032±0.0006 | +0.0027 ± 0.0006 |
| P$_8$ ~ 500 nm (Uniform wire) | -0.0077± 0.001 | **+0.0052± 0.001** | **+0.0001± 0.001** |

Since, major contribution of TO and LO phonon is due to ZB, the equation 2 and 3 are applied to E$_{2h}$ and TO, LO phonons, respectively. The calculated contribution of strain to the d$\omega$/dT$_{E2h, TO\ and\ LO}$ and are tabulated in table IV for a tapered MNW and a uniform wire. It can clearly be seen from table IV that d$\omega$/dT$_{E2h}$ due to strain increases as we go from base to tip, however for TO and LO phonons d$\omega$/dT$_{TO\ and\ LO}$ the same can be said but with much lesser certainty (marked in red/bold). The uncertainty can be correlated to lesser content of WZ. It is already noted that WZ content increases from base to tip and the trend of d$\omega$/dT contribution for these phonons also changes accordingly as given in table III.

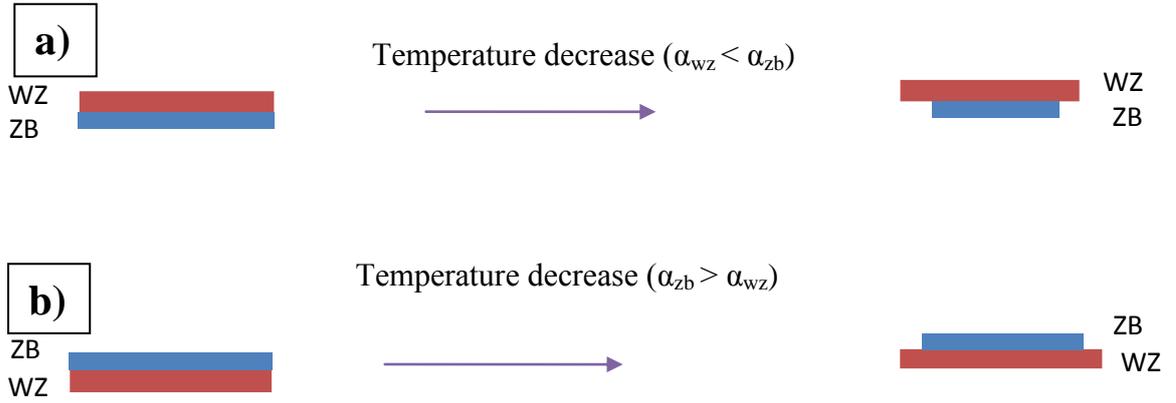

*FIG 12: Schematic diagram show a) the ZB as a substrate and WZ as a layer and b) the WZ as a substrate and ZB as a layer.*

The large error bars observed for some points in ω v/s T plots are mainly due to two reasons i) intensity of the TO/$E_{2h}$ is much smaller w.r.t. the other and ii) frequencies of the two are too close and hence, determination of position has larger uncertainty. The reason for smaller TO and $E_{2h}$ intensity is also two fold, i) smaller content of ZB/WZ phase and ii) changing of resonance itself due to different mixture of ZB to WZ phase [44]. Therefore, we concentrate on sign of $d\omega/dT_{E_{2h}, TO\ and\ LO}$ due to stress (Table IV), which is more definitive in nature and explain the same in the following. The in-plane thermal expansion of WZ ($\alpha_a$) is less compared to thermal expansion of ZB. As we decrease the temperature, the change in length of WZ is small compared to change in ZB (Fig. 12). In order to match atomic arrangement of ZB length (substrate layer), WZ will feel further compressive strain and therefore $E_{2h}$ (WZ) will blue shift on decrease in temperature giving rise to negative contribution to $d\omega/dT_{E_{2h}}$. Whereas, exactly opposite will happen for ZB on WZ, where mode frequency will red shift due to generation of tensile strain on decrease in temperature. This will give positive contribution of $d\omega/dT_{TO,LO}$. The negative and positive dω/dT for $E_{2h}$ and TO phonon, respectively can thus be explained as due to difference in thermal expansion of WZ and ZB phases of InAs. Using this temperature dependent Raman data, we estimate thermal expansion of WZ phase of InAs as described in the following.

The temperature dependent Raman shift, $\Delta\omega(T)$ at temperature T relative to the frequency $\omega_0$ at room temperature can be written as [53],

$$\Delta\omega(T) = \Delta\omega_{TE}(T) + \Delta\omega_S(T) + \Delta\omega_A(T) \qquad (4)$$

Where $\Delta\omega_{TE}(T)$, $\Delta\omega_S(T)$, $\Delta\omega_A(T)$ denote the frequency shift due to thermal expansion, interfacial strain and anharmonicity in lattice, respectively. In present case the range of temperature effect is taken to be below Debye temperature (250K) [42, 54] to avoid significant contribution due to anharmonicitiy. Therefore, we can neglect the anharmonic part. In an isotropic approximation, the term $\Delta\omega_{TE}(T)$ and $\Delta\omega_S(T)$ are given by [53,55] for WZ,

$$\Delta\omega_{TE}(T) = -\omega \left[ exp\left\{ -\gamma \int_0^T [\alpha_c(\tilde{T}) + 2\alpha_a(\tilde{T})] d\tilde{T} \right\} - 1 \right] \qquad (5)$$

$$\Delta\omega_S(T) = -\omega_0\, \gamma\, (2+\beta) \left[ (1+\varepsilon_g) \frac{1+\int_{T_g}^T \alpha_{zb}(T)dT}{1+\int_{T_g}^T \alpha_{wz}(T)dT} - 1 \right] \qquad (6)$$

Where $\alpha_c$ and $\alpha_a$ are the temperature dependent coefficient of linear thermal expansion parallel and perpendicular to the hexagonal c axis. $\varepsilon_g$ is average residual strain. Here, in our case this average residual strain is present due to mixed structure of ZB and WZ. The contribution of change in frequency due to change in temperature is expected due to interfacial strain. This interfacial strain in turn is arising due to difference in ZB and WZ thermal expansion, which is approximated to single WZ layer on ZB and vice versa, as we are dealing with average values of frequency, strain and WZ/ZB content over the scattering volume of Raman spectra. With this consideration, we can write $\Delta\omega_{TE}(T)$ for ZB [56] as,

$$\Delta\omega_{TE}(T) = \omega_0 \left[ exp\left\{ -3\gamma \int_0^T \alpha(T) d\tilde{T} \right\} - 1 \right] \qquad (7)$$

$$\Delta\omega_S(T) = -\omega_0 \gamma (2+\beta) \left[(1+\varepsilon_g) \frac{1+\int_{T_g}^{T} \alpha_{wz}(T)dT}{1+\int_{T_g}^{T} \alpha_{zb}(T)dT} - 1\right] \qquad (8)$$

$$\Delta\omega_S(T) = -\omega_0 \gamma (2+\beta) \left[(1+\varepsilon_g) \frac{1-\alpha_{wz}(T_g-T)}{1-\alpha_{zb}(T_g-T)} - 1\right] \qquad (9)$$

Growth temperature i.e. 698 K and $\varepsilon_g$ is average residual strain, which is now calculated from Raman spectra at 300 K. $\gamma$ is gruneisen parameter i.e. 1.21 for bulk InAs [57] which does not have strong temperature dependence and hence it is neglected for the region considered. The temperature dependent thermal expansion of ZB is taken from Sirota et al. which varies from 4.52 $*10^{-6}$ to 2.25 $*10^{-6}$/K for temperature variation from 300 K to 80 K [58]. $\omega_0$ is transverse optical phonon frequency, $\beta$ is ratio of elastic constant [59]. $\Delta\omega_S(T) = (\omega_T - \omega_{300}) - (T-300)*d\omega_{ZB}/dT$ is calculated from frequency at temperature in the range of 80 K- 190 K. $d\omega_{ZB}/dT$ for TO phonon is taken to be -0.01cm$^{-1}$/K [42]. Taking these parameters, we have calculated the effective thermal expansion coefficient of WZ for base, center and tip position i.e. $\Delta\omega_{IS}$ (ZB/WZ) is taken to be happening due to difference in thermal expansion of two phases. Here $\Delta\omega_{IS}$ (ZB/WZ) is taken to be the average value ZB/WZ phase. It varies with temperature from 190 K to 80 K for base ~ 13 to 10*10$^{-6}$ /K, center ~ 20 to 14*10$^{-6}$ /K and tip ~24 to 19*10$^{-6}$ /K. It is appropriate to note here that accuracy of these values is highest at 80 K, as TO phonon frequency shows lowest uncertainty i.e. 218±0.06 cm$^{-1}$, unlike 217±1.5 cm$^{-1}$ at 190 K. The effective in-plane thermal expansion of WZ is increasing from base to tip. This is consistent with the fact that compressive stress on WZ is reducing as we go from base to tip due to increase in percentage of WZ towards tip position, which increases the haxagonality and getting close to 2H structure. Thus, the in-plane lattice constant of WZ is increasing and accordingly the thermal expansion of WZ is increasing. However, one must remember that for this calculation, constant thermal expansion coefficient for ZB phase is taken, which may not be the case in reality. This approximation is expected to give overestimated values for thermal expansion coefficient for WZ.

With the above understanding, it may be interesting to revisit SRR spectra of bent MNWs. For smooth bent (Fig. 2(d)), from position 1 to 8, as we are go from larger (800 nm) to smaller (300 μm) diameter, it starts blue shifting right before the bend occurs and continues till it reaches the tip. More such data on smooth and sharp bent MNWs reveals that sharp bent is observed, wherever there is sudden change in diameter. The sudden change is diameter may lead to development of larger strain due to sudden and larger change in WZ/ZB ratio, thus giving rise to sharp bent. It is further interesting to note that sharp bent has probably lead to relaxation of the stress in the MNW and hence thereafter it shows red shift, as per change in diameter observed for tapered MNWs. Important to the contrast here that for smooth bent MNWs (Fig. 2(b)), instead continuous blue shift is observed for both the modes (TO and $E_{2h}$) even after smooth bent although diameter is decreasing there too. This suggests that smooth bent has not allowed complete relaxation of the stress generated. One more observation needs to be mentioned here, i.e. reduction in $E_{2h}$ and TO phonon frequencies going from $x(z,z)\bar{x}$ to $x(y,y)\bar{x}$ geometry as phonons are polarized in Z and Y directions may be due to biaxial strain in the growth direction. Both these aspects are being further investigated and will be published elsewhere.

**Conclusion**

To elucidate the frequency shift observed in TO phonon of InAs in large diameter (~ 800 nm), several uniform, tapered and bent InAs micro-nanowires (MNWs) are studied using spatially resolved Raman spectroscopy. Growth mapping studies of metal organic vapor phase epitaxial grown InAs NWs of diameter ~100-200 nm reported in literature suggest MNWs of large diameter (500 nm -2 μm) to have pure zinc blende (ZB) structure. However, spatially resolved unpolarized, polarized and wavelength dependent Raman spectroscopy has established presence of mixed phases in large diameter uniform and tapered MNWs. This is found to lead to strained ZB and wurtzite (WZ) structures present in these InAs MNWs. The strain was further studied using spatially resolved temperature dependent Raman spectroscopy. This shows that there is significant contribution of stress to temperature dependent behavior of phonon frequencies under these conditions and it has -ve and +ve contribution to $d\omega/dT$ for WZ and ZB phase, respectively. This can be explained using relative thermal expansion coefficients of ZB and WZ phases. Considering the heterostructure and knowing thermal expansion of ZB phases, we have calculated effective thermal expansion of WZ, which increases from base to tip, consistent with increase in lattice constant of WZ i.e. relaxation of compressive stress from base to tip. Further, it is found that as grown bent MNWs formation is associated with sudden change of diameter and consequently ZB to WZ content. This is being further investigated and will be published elsewhere.

## Acknowledgment

Authors acknowledge Dr. V. K. Dixit, Dr. S. D. Singh and Dr. T. K. Sharma for the growth of the sample. We also thank Dr. S. M. Oak and Dr. H. S. Rawat for their constant support during course of this work. Authors also acknowledge experimental support of Mr. Ajay Kumar Rathore for temperature dependent Raman measurements. VKG wishes to acknowledge help of Ms Ekta Rani in careful reading of the manuscript.

**Figure Captions**

FIG. 1: a) Optical image of uniform InAs MNW (dia ~ 1.2 μm) and b) SRR spectra at positions (2-5) as marked in the optical image. Inset of Fig. 1a shows Raman spectrum for position 1 marked in the optical image. Cumulative fit (red solid line) to the raw data (＋) and separate Lorentzian fits are shown with blue solid line.

FIG. 2: a) Optical image of sharp bent tapered MNW (base dia ~ 1.3 μm and tip ~ 400 nm, b) SRR spectra at positions (1-7) as marked in the optical image, c) optical image of smooth bent tapered MNW (base dia ~ 800 nm and tip ~ 300 nm) and d) SRRS at positions (1-8) as marked in the optical image. Cumulative fit (red solid line) to the raw data (＋) and separate Lorentzian fits are shown with blue solid line.

FIG. 3: a) Optical image of straight tapered MNW, wherein objective is focused near the base (dia ~ 2 μm), b) Optical image of same wire, wherein objective is focused at the tip (~ 400 nm) and c) SRRS at positions (1-6) as marked in both optical images. Cumulative fit (red solid line) to the raw data (＋) and separate Lorentzian fits are shown with blue solid line.

FIG. 4: a) Optical image of InAs micro sphere and b) SRRS at positions marked in Fig. 4a. Cumulative fit (red solid line) to the raw data (＋) and separate Lorentzian fits are shown with blue solid line.

FIG. 5: Schematic diagram of the Raman scattering configuration w.r.t. NW axis taken to be in Z direction for back scattering geometry. x and $\bar{x}$ axes are the incident and scattering polarization. Z axes is taken to be [111] and [0001] for ZB and WZ phases, respectively.

FIG. 6: Optical image of uniform wire (dia ~ 600 nm) a) in horizontal direction, b) in vertical direction and c) Polarized Raman data in x(z,z)$\bar{x}$ and x(y,y)$\bar{x}$ configuration at positions as marked in optical image (a) and (b), respectively. Cumulative fit (red solid line) to the raw data (＋) and separate Lorentzian compartments are shown with blue solid line.

FIG. 7: a) Optical image shows tapered MNW (base dia ~ 2 μm and center dia ~ 1.6 μm and tip ~ 800 nm), b) unpolarized Raman data at position marked for base, center and tip, c), d) and e) Polarized Raman data for base center and tip position, respectively in x(z,z)$\bar{x}$ and x(y,y)$\bar{x}$ configuration. Cumulative fit (red solid line) to the raw data (＋) and separate Lorentzian fits are shown with blue solid line.

FIG. 8: a) shows diameter v/s frequency for 6 Tapered wire (TW). Here plus (＋) sign with red, blue, green, black, gray, dark pink colors show TO of ZB for TW 1,2,3,4,5,6, respectively. Cross (X) with red, blue, green, black, gray, dark pink colors show $E_{2h}$ of WZ for TW 1,2,3,4,5,6, respectively and b) shows diameter v/s frequency for 3 uniform wires. Symbols (＋) and (X) show TO and $E_{2h}$, respectively.

FIG. 9: a) and b) Polarized Raman data performed with 488 nm and 442 nm excitation, respectively at same wire and at similar position in x(z,z)$\bar{x}$ and x(y,y)$\bar{x}$ configuration for uniform wire (dia ~ 1.2 μm).

FIG. 10: a), b) and c) show the temperature dependent Raman spectra in the temperature range 300K to 80K for base (~2 μm), center (~1.6 μm) and tip (~800 nm) position, respectively. Cumulative fit (red solid line) to the raw data (＋) and separate Lorentzian fits are shown with blue solid line. Frequency v/s temperature plot showing linear fit for TO, LO and $E_{2h}$ frequencies for d) base, e) center and f) tip position. The Error bar corresponds to the standard error to ω as obtained from nonlinear least square fir to the spectra. The solid line are the best fit to the data points, from which slope are calculated in fig. d)-f).

FIG. 11: a) Show the temperature dependent Raman spectra in the temperature range 300K to 80K for uniform wire (dia ~ 500 nm). Cumulative fit (red solid line) to the raw data (＋) and separate Lorentzian fits are shown with blue solid line and b) Frequency v/s temperature plot showing linear fit for TO, LO and $E_{2h}$ frequencies for uniform wire. The solid line are the best fit to the data points.

FIG. 12: Schematic diagram show a) the ZB as a substrate and WZ as a layer and b) the WZ as a substrate and ZB as a layer.

**Table captions:**

Table I: Raman selection rules for different scattering geometries are summarized.

Table II: Diameter of base, center and tip positions for tapered wires (TW) designated as 1, 2, 3, 4, 5 and 6 for which phonon frequencies are plotted in Fig. 8.

Table III: Frequencies of $E_{2h}$ ($\omega_l$) and TO ($\omega_h$), intensity ratio of TO to $E_{2h}$ phonons at 300K and 80K and $d\omega_{E2h}/dT$, $d\omega_{TO}/dT$, $d\omega_{LO}/dT$ for different positions on a tapered wires and for a uniform wire are summarized.

Table IV: strain component of $(d\omega_{E2h}/dT)_S$, $(d\omega_{TO}/dT)_S$ and $(d\omega_{LO}/dT)_S$ for different positions on a tapered wires and for a uniform wire are summarized.